\documentclass[pre,showpacs]{revtex4}
\usepackage{amssymb}
\usepackage[dvips]{graphicx}
\usepackage{amsmath}
\usepackage{graphicx}
\usepackage{makeidx}
\usepackage{subfigure}

\begin{document}

\title{Stabilization of one-dimensional solitons against the critical
collapse by quintic nonlinear lattices}
\author{Jianhua Zeng$^{1,2}$ and Boris A. Malomed$^{1}$}
\affiliation{$^{1}$Department of Physical Electronics, School of Electrical Engineering,
Faculty of Engineering, Tel Aviv University, Tel Aviv 69978, Israel\\
$^{2}$State Key Laboratory of Low Dimensional Quantum Physics, Department of
Physics, Tsinghua University, Beijing 100084, China}

\begin{abstract}
It has been recently discovered that stabilization of two-dimensional (2D)
solitons against the critical collapse in media with the cubic nonlinearity
by means of nonlinear lattices (NLs) is a challenging problem. We address
the 1D version of the problem, i.e., the nonlinear-Schr\"{o}dinger equation
(NLSE) with the quintic or cubic-quintic (CQ) terms, the coefficient in
front of which is periodically modulated in space. The models may be
realized in optics and Bose-Einstein condensates (BECs). Stability diagrams
for the solitons are produced by means of numerical methods and analytical
approximations. It is found that the sinusoidal NL stabilzes\emph{\ }%
solitons supported by the quintic-only nonlinearity in a narrow stripe in
the respective parameter plane, on the contrary to the case of the cubic
nonlinearity in 2D, where the stabilization of solitons by smooth spatial
modulations is not possible at all. The stability region is much broader in
the 1D CQ model, where higher-order solitons may be stable too.
\end{abstract}

\maketitle


\section{Introduction}

The use of periodic potentials, induced by photonic crystals or optical
lattices for controlling light beams in optical waveguides \cite{PhotCryst}
or matter waves in Bose-Einstein condensates (BECs) \cite{ref1} was the
subject of many works, see reviews \cite{ref2}-\cite{ref10}. It has been
demonstrated that lattice potentials balancing the self-focusing or
defocusing cubic nonlinearity uphold bright solitons---ordinary ones or gap
modes, respectively. The general topic of solitons in period potentials has
been reviewed in detail in Refs. \cite{ref2}-\cite{ref4}. In the
two-dimensional (2D) geometry, where the cubic ($\chi ^{(3)}$) self-focusing
leads to the critical collapse \cite{Berge'}, fundamental solitons in free
space (\textit{Townes solitons} \cite{Townes}) are unstable, representing a
separatrix between collapsing and decaying modes. Vortex solitons in the 2D
free space exist too \cite{Kruglov}, being subject to a still stronger
instability against azimuthal perturbations \cite{ref3,Ukraine}). However,
lattice potentials can stabilize 2D fundamental solitons and solitary
vortices alike \cite{2Dstabilization}. Moreover, the same potentials may
also stabilize 2D solitons against the \textit{supercritical} collapse,
driven by the self-focusing quintic ($\chi ^{(5)}$) term \cite{Radik}.

In 1D, the critical collapse is induced by the quintic nonlinearity, which
gives rise to the known 1D counterparts of the Townes solitons, see Eq. (\ref%
{exact}) below. The analogy to the 2D setting suggests a possibility of the
stabilization of the 1D variety of Townes solitons by means of the
single-well or lattice potentials, which was demonstrated in Refs. \cite%
{Sharp-soliton} and \cite{ref15}, respectively, with the help of numerical
simulations and a variational approximation (VA). As concerns physical
realizations, the quintic nonlinearity (acting, generally, in a combination
with the cubic term) was theoretically predicted \cite{quintic-theory} and
experimentally observed in diverse optical media, including liquids \cite%
{liquids,colloid}, glasses \cite{quintic-experiment} and ferroelectric films
\cite{ferroelectric}. Generally, this nonlinearity comes along with the
cubic losses. Nevertheless, the losses may be usually neglected over
relatively short propagation distances for which experiments are run. In
particular, the cubic-quintic (CQ, alias $\chi ^{(3)}:\chi ^{(5)}$) optical
nonlinearity may be efficiently adjusted in colloids formed of metallic
nanoparticles, by selecting the size of the particles and the colloidal
filling factor \cite{colloid}. Using these parameters, one may emulate a
medium with the quintic-only nonlinearity, i.e., $\chi ^{(3)}=0$, which is
the case especially interesting for the theoretical consideration, see
below. Another realization of an effectively quintic medium may be possible
in a composite material built as an alternation of narrow self-focusing and
defocusing layers \cite{Isaac}, in which the cubic $\chi ^{(3)}$ coefficient
is canceled on the average, while its $\chi ^{(5)}$ coefficient remains
uncompensated.

In terms of the BEC, the quintic term in the Gross-Pitaevskii equation (GPE)
may represent three-body interactions in a sufficiently dense condensate,
provided that the lossy part of these interactions may be neglected (the
latter depends on particular properties of the condensate, the use of the
Feshbach resonance for controlling the scattering length of inter-atomic
collisions, etc.) \cite{ref14}. On the other hand, a universal quintic term
appears in the effective GPE, in the absence of any three-body interactions,
as a result of the reduction of the dimension from 3 to 1 under the action
of the strong confinement in the transverse plane \cite{confinement,Lev}. In
the latter case, no dissipative nonlinearity appears, and the sign of the
effective quintic term is self-attractive, irrespective of the sign of the
scattering length which determines the self-attractive or repulsive sign of
the cubic nonlinearity. The relative strength of the respective quintic term
in comparison to the basic cubic one can be estimated as $\sim 5|a|n_{%
\mathrm{1D}}$ (see, e.g., Ref. \cite{Lev}), where $a$ is the scattering
length ($a<0$ corresponds to the attraction between atoms), and $n_{\mathrm{%
1D}}$ the 1D density of the condensate. For the typical situation with
matter-wave solitons containing $\sim 5000$ atoms in the area of size $\sim
2 $ $\mathrm{\mu }$m, with $a\sim -0.1$ nm \cite{Hulet}, the estimate yields
the value $\sim 1$, hence the quintic terms may be as important as the cubic
one.

More recently, the studies of the soliton dynamics in periodic potentials
were extended for \textit{nonlinear lattices }(NLs), which are represented
by \textit{pseudopotentials} \cite{ref9} induced by spatially periodic
modulations of the local strength of\ the nonlinearity, see original works
\cite{ref5}-\cite{2D} and recent review \cite{ref10}. In optics and BEC, NLs
may be engineered by means of various techniques, such as filling voids of
photonic crystals with properly selected materials, or using the Feshbach
resonance controlled by spatially patterned fields \cite{ref10} (in
particular, \textit{magnetic lattices} governing the local strength of the
Feshbach resonance have been built using sets of ferromagnetic films \cite%
{ferro}). In quasi-1D BECs, an NL can also be induced by periodically
modulating the transverse-confinement strength in the axial direction \cite%
{Luca}. Similarly, a periodic modulation of the local nonlinearity strength
in a planar optical waveguide may be created by varying the thickness of the
waveguiding layer. In particular, the periodically modulated quintic-only
and CQ nonlinearities, considered below, may be realized in the
above-mentioned colloid layers with the periodic variation of the transverse
thickness, or \ by applying a properly patterned magnetic field to the
colloidal suspension formed by ferromagnetic particles.

The previous analysis has demonstrated that NLs readily support stable 1D
solitons, but it is quite difficult, although possible, to stabilize 2D
solitons against the critical collapse by means of nonlinear
(pseudo)potentials. Namely, 2D sinusoidal lattices completely fail to
provide for the stabilization, while shapes with \emph{sharp edges}, such as
circles or stripes, are able to do it \cite{2DHS}-\cite{2D}. Although a
periodic lattice of nonlinear circles with sharp edges can support stable 2D
solitons, they are virtually the same as those supported by a lone circle,
i.e., the periodicity plays no essential role in the stabilization \cite%
{Vyslo}.

The results outlined above suggest to investigate the 1D counterpart of the
problem, i.e., stabilization of 1D solitons by means of NLs against the
respective critical collapse, accounted for by the $\chi ^{(5)}$
(quintic-only) or CQ nonlinearity. Below, we perform the analysis of the
model with the quintic-only and CQ self-focusing terms, whose strength is
subject to the periodic spatial modulation, by means of numerical methods
and analytical approximations. We find that a smooth sinusoidal $\chi ^{(5)}$
NL can stabilize 1D solitons, on the contrary to the negative result in 2D,
but only in a narrow parametric region. This finding is explained with the
help of an analytical approach. The sinusoidal CQ NL gives rise to a much
broader stability area than its quintic counterpart, which is demonstrated
by means of the VA, in addition to numerical results. A chain of nonlinear
domains with sharp edges may support stable solitons in 1D as well, but it
can be readily checked that, like their 2D counterparts, such solitons are
actually supported by lone nonlinear islands, the effect of the periodicity
being negligible.

\section{The model and analytical approximations}

In terms of optical realizations, the scaled form of the underlying
nonlinear Schr\"{o}dinger equation (NLSE) for the field amplitude, $u(x,z)$,
is
\begin{equation}
iu_{z}=-\frac{1}{2}u_{xx}-\left[ \varepsilon _{0}+\varepsilon _{1}\cos (2x)%
\right] |u|^{2}u-[g_{0}+g_{1}\cos (2x)]|u|^{4}u,  \label{GPE}
\end{equation}%
where $\varepsilon _{0}$ and $g_{0}$ are strengths of the uniform CQ terms, $%
\varepsilon _{1}$ and $g_{1}$ represent the NL, and the scale of coordinate $%
x$ is fixed by setting the NL period to be $\pi $. Below, the remaining
scaling invariance is used to fix $g_{1}\equiv 1$. The corresponding GPE for
the mean-field wave function of BEC is obtained from Eq. (\ref{GPE}) by
replacing propagation distance $z$ with time $t$. A model featuring the
periodic modulation of the cubic nonlinearity along the evolution variable ($%
z$), rather than along $x$, was studied too, supporting stable motion of
solitons in the discrete version of the 1D model \cite{Cuevas}.

Stationary solutions to Eq. (\ref{GPE}) with propagation constant $k$ are
sought for as $u(x,z)=U(x)\exp (ikz)$, where real function $U(x)$ satisfies
equation
\begin{equation}
kU=\frac{1}{2}U_{xx}+\left[ \varepsilon _{0}+\varepsilon _{1}\cos (2x)\right]
U^{3}+[g_{0}+g_{1}\cos (2x)]U^{5},  \label{U}
\end{equation}%
which can be derived from the Lagrangian,

\begin{eqnarray}
L &=&\frac{1}{2}\int_{-\infty }^{+\infty }\left\{ kU^{2}+\frac{1}{2}%
U_{x}^{2}-\frac{1}{2}\left[ \varepsilon _{0}+\varepsilon _{1}\cos (2x)\right]
U^{4}\right.   \notag \\
&&\left. -\frac{1}{3}\left[ g_{0}+g_{1}\cos (2x)\right] U^{6}\right\} dx.
\label{L}
\end{eqnarray}%
In the case of $\varepsilon _{0,1}=g_{1}=0$, the above-mentioned unstable
Townes-type solution of the quintic NLSE with $g_{0}>0$ is well known:%
\begin{equation}
U^{2}\left( x\right) =\sqrt{3k/g_{0}}\mathrm{sech}\left( \sqrt{8k}x\right) ,
\label{exact}
\end{equation}%
whose norm (the total power or number of atoms, in terms optics or BEC,
respectively) does not depend on $k$,
\begin{equation}
N=\int_{-\infty }^{+\infty }U^{2}\left( x\right) dx=\frac{\pi }{2}\sqrt{%
\frac{3}{2g_{0}}},  \label{Townes}
\end{equation}%
which is the characteristic feature of Townes solitons \cite{Berge'}. The
addition of an arbitrarily weak cubic self-focusing term, with constant
coefficient $\varepsilon _{0}>0$, lifts the degeneracy and turns solution (%
\ref{exact}) into exact \emph{stable} solutions of the CQ equation \cite%
{Bulgaria,Peli}. The addition of the self-defocusing cubic term with $%
\varepsilon _{0}<0$ lifts the degeneracy too, but makes the entire soliton
family unstable \cite{Peli,Lev}.

To apply the VA \cite{VA}, we adopt the Gaussian ansatz for the soliton with
width $W$ and norm $N$:
\begin{equation*}
U^{2}\left( x\right) =\frac{N}{\sqrt{\pi }W}\exp \left( -\frac{x^{2}}{W^{2}}%
\right) .
\end{equation*}%
The substitution of the ansatz into Lagrangian (\ref{L}) yields the
effective Lagrangian,
\begin{equation*}
L_{\mathrm{eff}}=\frac{N}{2}\left[ k+\frac{1}{4W^{2}}-\frac{N}{2\sqrt{2\pi }W%
}\left( \varepsilon _{0}+\varepsilon _{1}e^{-\frac{W^{2}}{2}}\right) -\frac{%
N^{2}}{3\sqrt{3}\pi W^{2}}(g_{0}+g_{1}e^{-\frac{W^{2}}{3}})\right] ,
\end{equation*}%
which gives rise to the corresponding variational equations, $\partial L_{%
\mathrm{eff}}/\partial W=\partial L_{\mathrm{eff}}/\partial N=0$:
\begin{equation}
\frac{-1}{4W^{2}}+\frac{N\left( \varepsilon _{0}+\varepsilon _{1}e^{-\frac{%
W^{2}}{2}}\right) }{\sqrt{2\pi }W}+\frac{N^{2}\left( g_{0}+g_{1}e^{-\frac{%
W^{2}}{3}}\right) }{\sqrt{3}\pi W^{2}}=k,  \label{W}
\end{equation}%
\begin{gather}
\frac{-1}{2W^{2}}+\frac{N}{2\sqrt{2\pi }}\left[ \frac{\varepsilon _{0}}{W}%
+\varepsilon _{1}\left( \frac{1}{W^{2}}+W\right) e^{-\frac{W^{2}}{2}}\right]
\notag \\
+\frac{2N^{2}}{3\sqrt{3}\pi }\left[ \frac{g_{0}}{W^{2}}+g_{1}\left( \frac{1}{%
W^{2}}+\frac{1}{3}\right) e^{-\frac{W^{2}}{3}}\right] =0.  \label{N}
\end{gather}%
As shown below, the VA produces reasonable results in the model with the CQ
nonlinearity. The VA may be extended to investigate the stability of
perturbed (nonstationary) solitons by including the chirp into the ansatz,
but this is not really necessary, as the stability of the solitons which are
reasonably well approximated by the ansatz is correctly predicted by the VK
criterion, as shown below.

The VA fails to produce accurate results for the quintic-only model ($%
\varepsilon _{0,1}=0$), unlike the CQ one (see below). Nevertheless, in the
quintic model another analytical approximation can be applied to broad
solitons with $k\lesssim 1$ and $g_{0}\gg g_{1}$. In this case, an
approximate correction to soliton (\ref{exact}), generated by term $\sim
g_{1}$ in Eq. (\ref{U}), is found to be
\begin{equation}
U_{1}\approx \left( g_{1}/2\right) \left( 3k/g_{0}\right) ^{5/4}\left[
\mathrm{sech}\left( \sqrt{8k}x\right) \right] ^{5/2}\cos \left( 2x\right) ,
\label{U1}
\end{equation}%
and the corresponding correction to norm (\ref{Townes}), which lifts its
degeneracy, is
\begin{equation}
N_{1}\approx \left( 3\sqrt{3}\pi /4\right) g_{1}g_{0}^{-3/2}\exp \left( -\pi
/\left( 2\sqrt{2k}\right) \right) .  \label{N1}
\end{equation}%
Thus, correction (\ref{U1}) replaces the single value (\ref{Townes}) by the
exponentially narrow stripe of values (\ref{N1}), within which the norm
satisfies the \textit{Vakhitov-Kolokolov} (VK) necessary stability
criterion, $dN/dk>0$ \cite{VK,Berge'}, suggesting that there may exist a
narrow stability stripe for the 1D solitons in the quintic-NL model. This
expectation is confirmed by numerical results presented below.

For the sake of the comparison of the models with the critical (quintic and
cubic, respectively) nonlinearities in 1D and 2D, it is relevant to mention
that an approximation similar to that based on Eq. (\ref{U1}) can be
developed too for broad 2D solitons in the $\chi ^{(3)}$ model with the full
2D or quasi-1D sinusoidal modulation of the nonlinearity, making use of
formula $\int_{0}^{2\pi }\cos \left( 2r\cos \theta \right) d\theta =2\pi
J_{0}\left( 2r\right) $ for the angular integration in the calculation of
the ensuing correction for the norm, and of the asymptotic form for the 2D
Townes soliton, $U_{0}(r)\sim r^{-1/2}\exp \left( -\sqrt{2k}r\right) $.
However, the eventual result for the 2D setting turns out to be opposite: $%
dN^{(\mathrm{2D})}/dk<0$, which explains the failure to of the sinusoidal NL
to stabilize the 2D solitons.

\section{Localized modes in the quintic model}

Stationary solutions to Eq. (\ref{GPE}), i.e., solutions of Eq. (\ref{U}),
were found by means of the imaginary-time-propagation method \cite{ref16},
and their stability was tested by subsequent simulations of the real
propagation.

\begin{figure}[tbp]
\begin{center}
\includegraphics[height=4.cm]{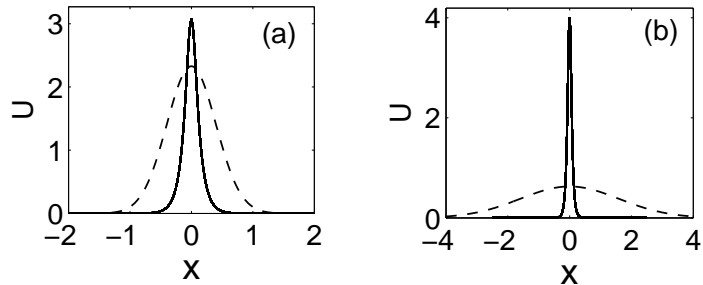}
\end{center}
\caption{Shapes of stable (a) and unstable (b) solitons in the quintic model
with $\protect\varepsilon =0,~g_{0}=0.5$ are shown for $k=69.8$, $N=1.65$,
and $k=71.5$, $N=1.78$, respectively. The dashed lines represent the shapes
predicted by the variational approximation. }
\label{Fig1}
\end{figure}

\begin{figure}[tbp]
\begin{center}
\includegraphics[height=5.cm]{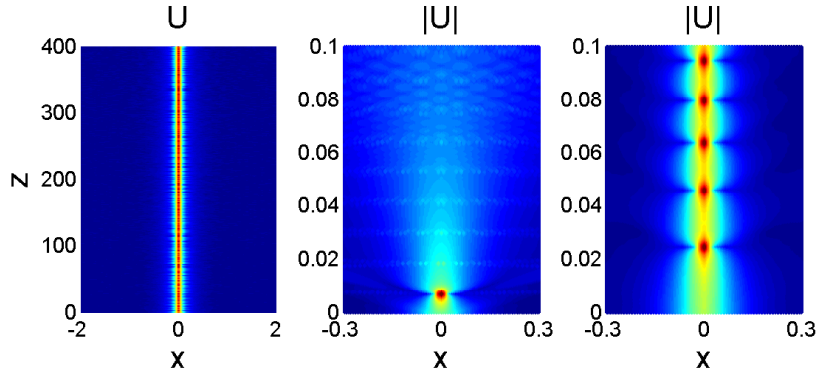}
\end{center}
\caption{(Color online) The evolution of the stable and unstable solitons
from Fig. \protect\ref{Fig1} is displayed in the left and central panels,
respectively. In addition, the right panel shows an example of the unstable
evolution towards the collapse in the quintic model (unlike the decay in the
central panel), for the soliton with $k=88.2$ and $N=1.85$.}
\label{Fig2}
\end{figure}

Examples of the shape and evolution of stable and unstable solitons
supported by the quintic-only ($\varepsilon _{0,1}=0$) NL are displayed in
Figs. \ref{Fig1} and \ref{Fig2}. In Fig. \ref{Fig1}, the comparison with the
numerical findings demonstrates that the Gaussian ansatz fails to correctly
predict the actual shape of the solitons in the case of the quintic
nonlinearity, therefore only numerical results are reported in this section.
For very narrow solitons, whose width is much smaller than the NL period
[such as the soliton shown in Fig. \ref{Fig1}(b)], Eq. (\ref{U}) is
virtually tantamount to the equation with the constant nonlinearity
coefficient, $\tilde{g}_{0}\equiv g_{0}+g_{1}=g_{0}+1$ (recall we fix $%
g_{1}\equiv 1$), and, accordingly, the narrow solitons may be considered as
Townes modes (\ref{exact}), with $g_{0}$ replaced by $\tilde{g}_{0}$.
Therefore, the narrow solitons are unstable, see the central and right
panels in Fig. \ref{Fig2}. The instability manifests itself as a quick decay
of the soliton (see the central panel in Fig. \ref{Fig2}), or as the
evolution towards the collapse, see the right panel in Fig. \ref{Fig2}. Note
the great difference in the scale of $z$ in the panels of Fig. \ref{Fig2}
demonstrating the stable and unstable evolution (i.e., the stable behavior
persists indefinitely long, while the instability sets in quickly). It is
relevant to compare these scales with the diffraction length corresponding
to the solitons displayed in Fig. \ref{Fig1}, which is $z_{\mathrm{diffr}%
}\sim 0.1$. On the other hand, it is relevant to stress that the GPE in the
form of Eq. (\ref{GPE}) may predict only the onset of the collapse, while
its further development requires a more accurate (microscopic) description,
in terms of the Hartree-Fock-Bogoliubov equations \cite{collapse}.

\begin{figure}[tbp]
\begin{center}
\includegraphics[height=4.cm]{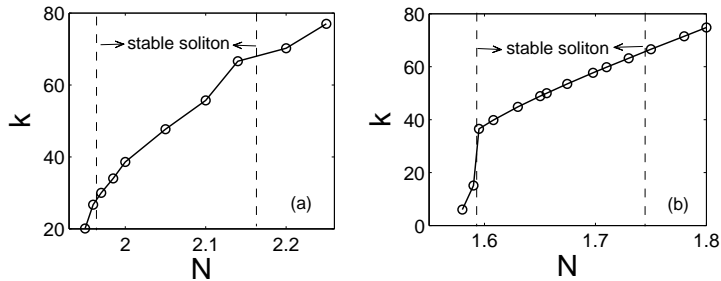}
\end{center}
\caption{The propagation constant vs. the norm for the soliton families in
the quintic model ($\protect\varepsilon _{0,1}=0$) for $g_{0}=0$ (a) and $%
g_{0}=0.5$ (b). The designated stability intervals were identified via
direct simulations.}
\label{Fig3}
\end{figure}

The numerically found dependence $k(N)$ for the solitons in the quintic
model is displayed in Fig. \ref{Fig3} for two different values of $g_{0}$.
While these curves obey the VK criterion, direct simulations of the
perturbed evolution of the solitons demonstrate that the solitons are stable
only in narrow intervals of the values of $N$ (recall that the VK criterion
is only necessary, but not sufficient, for the stability).

\begin{figure}[tbp]
\begin{center}
\includegraphics[height=4.cm]{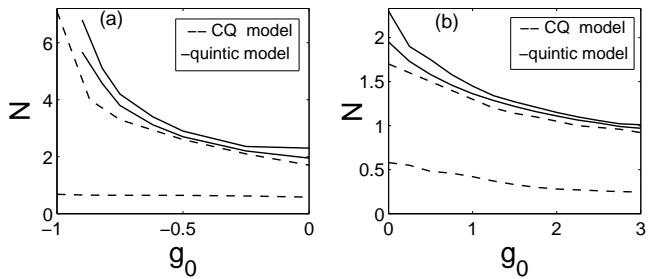}
\end{center}
\caption{Stability regions (located between the borders) for the solitons in
the parameter plane of the quintic and CQ models (solid and dashed borders,
respectively), shown separately for the self-defocusing (a) and
self-focusing (b) sign of the constant part of the quintic coefficient, $%
g_{0}$.}
\label{Fig4}
\end{figure}

The numerically found stability region of solitons in the quintic model is
located between the solid borders in Fig. \ref{Fig4}, for both positive and
negative (self-focusing and defocusing) values of the constant quintic
coefficient, $g_{0}\gtrless 0$. The overall shape of the narrow region is
quite close to
\begin{equation}
N=\left( \pi /2\right) \sqrt{3/\left[ 2\left( 1+g_{0}\right) \right] },
\label{Ncr}
\end{equation}%
which is explained by the fact that the solitons are always found near
values of $N$ corresponding to the critical one, given by Eq. (\ref{Townes}%
), with $g_{0}$ substituted by the above-mentioned local value, $\tilde{g}%
_{0}\equiv g_{0}+1$. This argument also explains nonexistence of solitons at
$g_{0}\leq -1$. The shrinkage of the stability stripe at large positive $%
g_{0}$ is a manifestation of the relative weakness of the NL in this limit,
and is consistent with the exponential weakness of the NL-induced
stabilization predicted by the above analysis, see Eq. (\ref{N1}). Above the
narrow stripe, the unstable solitons suffer the collapse, while below the
stripe they decay, which may be realized as the \textit{delocalization
transition}, similar to that studied in systems with linear lattice
potentials \cite{2Dstabilization,deloc}.

\section{Localized modes in the CQ (cubic-quintic) model}

\begin{figure}[tbp]
\begin{center}
\includegraphics[height=4.cm]{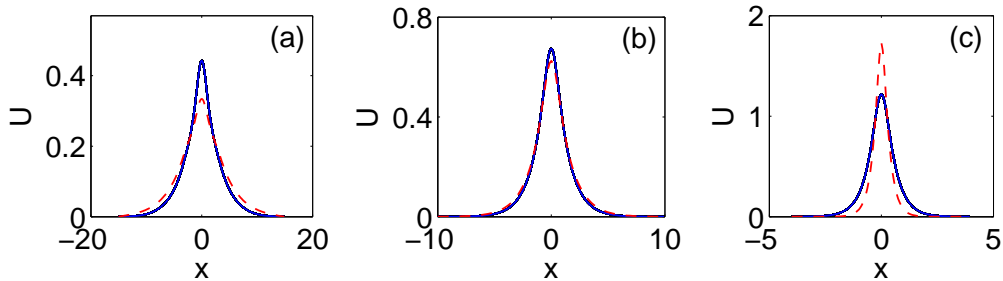}
\end{center}
\caption{(Color online) Typical examples of broad stable solitons in the CQ
model with $g_{0}=0.25$, $\protect\varepsilon _{0}=\protect\varepsilon %
_{1}=g_{1}=1$: (a) $k=0.26$, $N=0.65$; (b) $k=0.57$, $N=0.8$; (c) $k=3.2$, $%
N=1.2$. The dashed lines represent shapes predicted by the variational
approximation for the same values of the parameters.}
\label{Fig5}
\end{figure}

\begin{figure}[tbp]
\begin{center}
\includegraphics[height=4.cm]{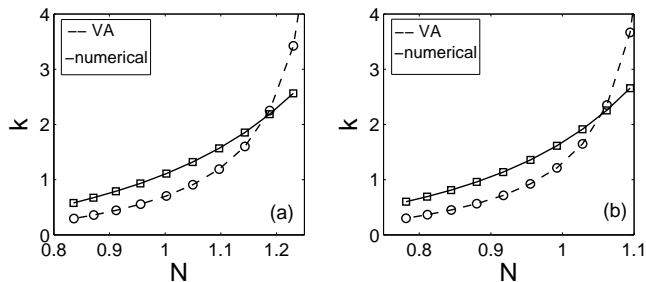}
\end{center}
\caption{Comparison of the variationally predicted and numerically found
dependences $k(N)$ in the CQ model, with $\protect\varepsilon _{0}=\protect%
\varepsilon _{1}=g_{1}=1$, for $g_{0}=0$ (a) and $g_{0}=0.5$ (b) .}
\label{Fig6}
\end{figure}

\begin{figure}[tbp]
\begin{center}
\includegraphics[height=4.cm]{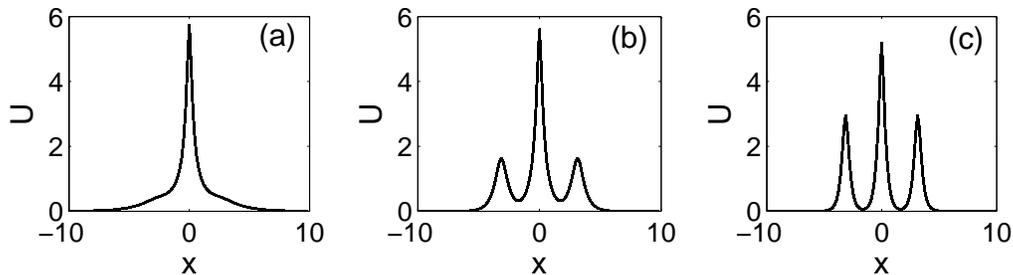}
\end{center}
\caption{Examples of a stable broad soliton (a) and stable solitons with
symmetric side peaks (b, c) in the CQ model with $\protect\varepsilon _{0}=%
\protect\varepsilon _{1}=g_{1}=1$ and $g_{0}=-1.05$: (a) $k=2.88$, $N=19.5$;
(b) $k=5.22$, $N=21.3$; (c) $k=9.23$, $N=22$.}
\label{Fig7}
\end{figure}

Unlike the model with the quintic-only nonlinearity, the one with the CQ
terms gives rise to broad solitons, as shown in Fig. \ref{Fig5} [see also
Fig. \ref{Fig7}(a) below]. Due to this fact, the predictions produced by the
VA for the CQ model are in reasonable agreement with the numerical findings,
as can be seen in Figs. \ref{Fig5} and \ref{Fig6}. On the other hand, the
agreement deteriorates at large values of $N$, when the quintic term
dominates, driving the system to the regime considered in the previous
section.

Dependences $k(N)$ in Fig. \ref{Fig6} satisfy the VK criterion, and
simulations demonstrate that the solitons in the CQ model, in contrast to
the quintic-only one, are always stable when they obey the criterion.
Accordingly, the stability area found in the parameter plane of the CQ
system, which is located between the dashed borders in Fig. \ref{Fig4}, is
much larger than its counterpart in the quintic model. The upper boundary of
the area, which is determined by the collapse driven by the local quintic
nonlinearity, is approximated by Eq. (\ref{Ncr}).

\begin{figure}[tbp]
\begin{center}
\includegraphics[height=4.cm]{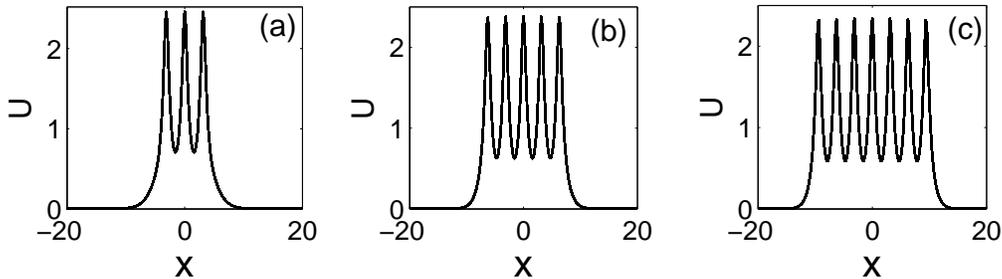}
\end{center}
\caption{Examples of stable in-phase bound states of three (a), five (b),
and seven (c) solitons in the CQ model with $\protect\varepsilon _{0}=%
\protect\varepsilon _{1}=g_{1}=1$, and $g_{0}=-1.25$: (a) $k=0.73$, $N=20$;
(b) $k=1.12$, $N=30$; (c)$k=0.73$, $N=40$. }
\label{Fig8}
\end{figure}

Another essential difference from the quintic-only model is that the
stability area in the CQ system extends to $g_{0}<-1$, where the quintic
nonlinearity is completely defocusing, but the solitons can be supported by
the cubic term. In this case, stable solitons develop side peaks, as shown
in Fig. \ref{Fig7}, and form odd-numbered \emph{stable} in-phase bound
states, see examples of three-, five- and seven-soliton complexes in Fig. %
\ref{Fig8}. The transition from the higher-order soliton, featuring the side
peaks, to the bound complexes occurs when the height of the additional peaks
becomes equal to that of the central maximum. Bound states of three in-phase
solitons were previously found in the 1D cubic-NL\ model, but they were
completely unstable \cite{ref5}.

\begin{figure}[tbp]
\begin{center}
\includegraphics[height=4.cm]{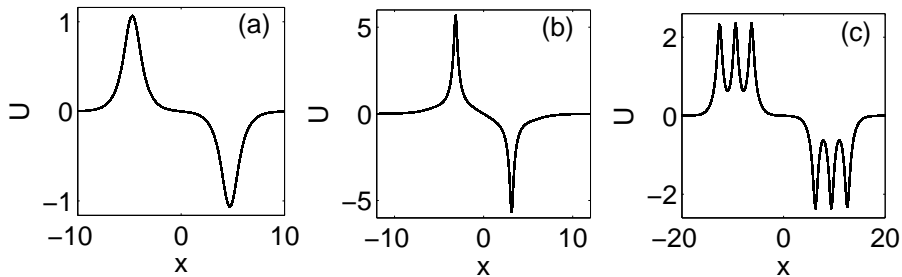}
\end{center}
\caption{Examples of stable dipole solitons (a, b), and of a stable cluster
dipole (c), in the CQ model with $\protect\varepsilon _{0}=\protect%
\varepsilon _{1}=1$. Parameters are: (a) $g_{0}=0.5$, $k=1.1$, $N=3.4$; (b) $%
g_{0}=-1.05$, $k=2$, $N=38$; (c) $g_{0}=-1.25$, $k=1.1$, $N=36$. Note that
the quintic term is self-defocusing in cases (b) and (c), but self-focusing,
on the average, in (a). }
\label{Fig9}
\end{figure}

The present system also supports stable dipoles, i.e., bound states of two
solitons with opposite signs, as well as stable dipolar complexes built of
two multi-peak soliton clusters, as shown in Fig. \ref{Fig9}. The simplest
two-peak dipoles are similar to their counterparts found in the cubic-NL
model \cite{ref5}.

\section{Conclusions}

This work aims to introduce the smooth sinusoidal NL (nonlinear lattice)
acting on the critical quintic ($\chi ^{(5)}$) nonlinearity in the 1D
setting, as well as the system with the CQ (cubic-quintic) NL. The models
approximate settings which are relevant to optics and BEC. Stability regions
for solitons in these systems have been identified, being narrow in the
quintic-only model, and much broader in the CQ one. In the latter case, the
results may be predicted analytically by means of the VA (variational
approximation), in the combination with the VK stability criterion. The
narrow stability stripe in the quintic model, and its absence in the
recently studied 2D models with sinusoidal NLs acting on the cubic
nonlinearity (which is critical in the 2D geometry) were also qualitatively
explained by means of an analytical approximation. A notable feature of the
CQ NL is that it generates stable broad solitons, which may develop side
peaks and form stable bound states, both in-phase and out-of-phase ones.

It may be interesting to study other dynamical features in models of this
type, such as mobility of the solitons. A relevant extension should also be
consideration of two-component systems similar to the scalar models
introduced in this work, cf. the analysis of two-component solitons in the
cubic-NL system, which was reported in Ref. \cite{Barcelona-vector}.

\emph{Acknowledgments}. The work of J.Z. was supported by a postdoctoral
fellowship provided by the Tel Aviv University, and by grant No. 149/2006
from the German-Israel Foundation.


\begin{thebibliography}{99}
\bibitem{PhotCryst} J. D. Joannopoulos, S. G. Johnson, J. N. Winn, and R. D.
Meade, \textit{Photonic Crystals: Molding the Flow of Light} (Princeton
University Press: Princeton, 2008).

\bibitem{ref1} C. J. Pethick and H. Smith, \textit{Bose-Einstein condensate
in dilute gas} (Cambridge University Press: Cambridge, 2008).

\bibitem{ref2} V. A. Brazhnyi, and V. V. Konotop, Mod. Phys. Lett. B \textbf{%
18}, 627 (2004); O. Morsch and M. Oberthaler, Rev. Mod. Phys. \textbf{78},
179 (2006).

\bibitem{ref3} B. A. Malomed, D. Mihalache, F. Wise, and L. Torner, J.
Optics B: Quant. Semicl. Opt. \textbf{7}, R53 (2005).

\bibitem{discrete-optics} F. Lederer, G. I. Stegeman, D. N. Christodoulides,
G. Assanto, M. Segev, and Y. Silberberg, Phys. Rep. \textbf{463}, 1 (2008).

\bibitem{ref4} Y. V. Kartashov, V. A. Vysloukh, and L. Torner, Progress in
Optics \textbf{52}, 63 (ed. by E. Wolf: North Holland, Amsterdam, 2009).

\bibitem{ref10} Y. V. Kartashov, B. A. Malomed, and L. Torner, Rev. Mod.
Phys. \textbf{83}, 247 (2011).

\bibitem{Berge'} L. Berg\'{e}, Phys. Rep. \textbf{303}, 259 (1998).

\bibitem{Townes} R. Y. Chiao, E. Garmire, and C. H. Townes, Phys. Rev. Lett.
\textbf{13}, 479 (1964).

\bibitem{Kruglov} V. I. Kruglov, Y. A. Logvin, and V. M. Volkov, J. Mod.
Opt. \textbf{39}, 2277 (1992).

\bibitem{Ukraine} A. I. Yakimenko, Yu. A. Zaliznyak, and Yu. S. Kivshar,
Phys. Rev. E \textbf{71}, 065603(R) (2005).

\bibitem{2Dstabilization} B. B. Baizakov, B. A. Malomed, and M. Salerno,
Europhys. Lett. 63, 642 (2003); Phys. Rev. A \textbf{70}, 053613 (2004); J.
Yang and Z. H. Musslimani, Opt. Lett. \textbf{28}, 2094 (2003).

\bibitem{Radik} R. Driben and B. A. Malomed, Eur. Phys. J. D \textbf{50},
317 (2008).

\bibitem{Sharp-soliton} Yu. B. Gaididei, J. Schjodt-Eriksen, and P. L.
Christiansen, Phys. Rev. E \textbf{60}, 4877 (1999).

\bibitem{ref15} F. Kh. Abdullaev and M. Salerno, Phys. Rev. A \textbf{72},
033617 (2005); G. L. Alfimov, V. V. Konotop, and P. Pacciani, Phys. Rev. A
\textbf{75}, 023624 (2007).

\bibitem{quintic-theory} G. S. Agarwal and S. Dutta Gupta, Phys. Rev. A
\textbf{38}, 5678 (1988); K. Dolgaleva, R. W. Boyd, J. E. Sipe, Phys. Rev. A
\textbf{76}, 063806 (2007).

\bibitem{liquids} C. Zhan, D. Zhang, D. Zhu, D. Wang, Y. Li, D. Li, Z. Lu,
L. Zhao, and Y. Nie, J. Opt. Soc. Am. B \textbf{19}, 369 (2002); R. A.
Ganeev, M. Baba, M. Morita, A. I. Ryasnyansky, M. Suzuki, M. Turu, H.
Kuroda, J. Opt. A: Pure Appl. Opt. \textbf{6}, 282 (2004).

\bibitem{colloid} E. L. Falc\~{a}o-Filho, C. B. de Ara\'{u}jo, and J. J.
Rodrigues Jr., J. Opt. Soc. Am. B \textbf{24}, 2948 (2007).

\bibitem{quintic-experiment} C. Zhan, D. Zhang, D. Zhu, D. Wang, Y. Li, D.
Li, Z. Lu, L. Zhao, and Y. Nie, J. Opt. Soc. Am. B \textbf{19}, 369 (2002);
G. Boudebs, S. Cherukulappurath, H. Leblond, J. Troles, F. Smektala, and F.
Sanchez, Opt. Commun. 219, 427 (2003); F. Smektala, C. Quemard, V. Couderc,
and A. Barth\'{e}l\'{e}my, J. Non-Cryst. Solids \textbf{274}, 232 2000; K.
Ogusu, J. Yamasaki, S. Maeda, M. Kitao, and M. Minakata, Opt. Lett. \textbf{%
29}, 265 (2004); F. Sanchez, G. Boudebs, S. Cherukulappurath, H. Leblond, J.
Troles, and F. Smektala, J. Nonlinear Opt. Phys. Mater. \textbf{13}, 7
(2004).

\bibitem{ferroelectric} B. Gu, Y. Wang, W. Ji, and J. Wang, Appl. Phys.
Lett. \textbf{95}, 041114 (2009).

\bibitem{Isaac} I. Towers and B. A. Malomed, J. Opt. Soc. Am. \textbf{19},
537 (2002).

\bibitem{ref14} P. F. Bedaque, E. Braaten, and H. W. Hammer, Phys. Rev.
Lett. \textbf{85}, 908 (2000); A. Gammal, T. Frederico, L. Tomio, and F. K.
Abdullaev, Phys. Lett. A \textbf{267}, 305 (2000); F. Kh. Abdullaev, A.
Gammal, L. Tomio, and T. Frederico, Phys. Rev. A \textbf{63}, 043604 (2001);
E. Braaten, H. W. Hammer, and T. Mehan, Phys. Rev. Lett. \textbf{88}, 040401
(2002); J. W. Michael, \textit{ibid}. \textbf{89}, 140402 (2002); W. Zhang,
E. M. Wright, H. Pu, and P. Meystre, Phys. Rev. A \textbf{68}, 023605
(2003); E. Fersino, G. Mussardo, and A. Trombettoni, \textit{ibid}. \textbf{%
77}, 053608 (2008).

\bibitem{confinement} A. E. Muryshev, G. V. Shlyapnikov, W. Ertmer, K.
Sengstock, and M. Lewenstein, Phys. Rev. Lett. \textbf{89}, 110401 (2002);
L. Salasnich, A. Parola, and L. Reatto, Phys. Rev. A \textbf{65}, 043614
(2002); \textit{ibid}. \textbf{66}, 043603 (2002); L. D. Carr and J. Brand,
Phys. Rev. Lett. \textbf{92}, 040401 (2004); Phys. Rev. A \textbf{70},
033607 (2004).

\bibitem{Lev} L. Khaykovich and B. A. Malomed, Phys. Rev. A \textbf{74},
023607 (2006).

\bibitem{Hulet} K. E. Strecker, G. B. Partridge, A. G. Truscott, and R. G.
Hulet, New J. Phys. \textbf{5}, 73.1 (2003).

\bibitem{ref9} T. Mayteevarunyoo, B. A. Malomed, and G. Dong, Phys. Rev. A
\textbf{78}, 053601 (2008).

\bibitem{ref5} H. Sakaguchi and B. A. Malomed, Phys. Rev. E \textbf{72}%
,046610 (2005).

\bibitem{ref5b} G. Theocharis, P. Schmelcher, P. G. Kevrekidis, and D. J.
Frantzeskakis, Phys. Rev. E \textbf{72}, 033614; F. Kh. Abdullaev and J.
Garnier, \textit{ibid}. \textbf{72},061605(R) (2005); Y. Sivan, G. Fibich,
and M. I. Weinstein, Phys. Rev. Lett. \textbf{97}, 193902 (2006); Z. Rapti,
P. G. Kevrekidis, V. V. Konotop, and C. K. R. T. Jones, J. Phys. A \textbf{40%
}, 14151 (2007).

\bibitem{ferro} S. Ghanbari, T. D. Kieu, A. Sidorov, and P. Hannaford, J.
Phys. B: At. Mol. Opt. Phys. \textbf{39}, 847 (2006); A. Abdelrahman, P.
Hannaford, and K. Alameh, Opt. Exp. \textbf{17}, 24358 (2009).

\bibitem{Luca} L. Salasnich, A. Cetoli, B. A. Malomed, F. Toigo, and L.
Reatto, Phys. Rev. A \textbf{76}, 013623 (2007).

\bibitem{2DHS} H. Sakaguchi and B. A. Malomed, Phys. Rev. E \textbf{73},
026601 (2006).

\bibitem{Vyslo} Y. V. Kartashov, B. A. Malomed, V. A. Vysloukh, and L.
Torner, Opt. Lett. \textbf{34}, 770 (2009).

\bibitem{2D} Y. Sivan, G. Fibich, B. Ilan, and M. I. Weinstein, Phys. Rev. E
\textbf{78}, 046602 (2008); N. V. Hung, P. Zi\'{n}, M. Trippenbach, and B.
A. Malomed, \textit{ibid}. \textbf{82}, 046602 (2010); T. Mayteevarunyoo, B.
A. Malomed, and A. Reoksabutr, J. Mod. Opt. \textbf{58}, 1977 (2011).

\bibitem{Cuevas} J. Cuevas, B. A. Malomed and P. G. Kevrekidis, Phys. Rev. E
\textbf{71}, 066614 (2005).

\bibitem{Bulgaria} K. I. Pushkarov, D. I. Pushkarov, and I. V. Tomov, Opt.
Quantum Electron. \textbf{11}, 471 (1979).

\bibitem{Peli} D. E. Pelinovsky, Y. S. Kivshar and V. V. Afanasjev, Physica
D 116, 121 (1998).

\bibitem{VA} B. A. Malomed, Progress in Optics \textbf{43}, 71 (ed. by E.
Wolf: North Holland, Amsterdam, 2002).

\bibitem{VK} M. Vakhitov and A. Kolokolov, Radiophys. Quantum. Electron.
\textbf{16}, 783 (1973).

\bibitem{ref16} W. Bao and Q. Du, SIAM J. Sci. Comput. \textbf{25}, 1674
(2004).

\bibitem{collapse} S. Wuster, B. J. Dabrowska-Wuster, A. S. Bradley, M. J.
Davis, P. B. Blakie, J. J. Hope, and C. M. Savage, Phys. Rev. A \textbf{75},
043611 (2007).

\bibitem{deloc} B. B. Baizakov and M. Salerno, Phys. Rev. A \textbf{69},
013602 (2004).

\bibitem{Barcelona-vector} Y. V. Kartashov, B. A. Malomed, V. A. Vysloukh,
and L. Torner, Opt. Lett. \textbf{34}, 3625 (2009).
\end{thebibliography}
\end{document}